\documentstyle[12pt]{article}

\newcommand{\be}{\begin{equation}}\newcommand{\ee}{\end{equation}}
\newcommand{\bea}{\begin{eqnarray}}\newcommand{\eea}{\end{eqnarray}}
\newcommand{\ba}{\begin{array}{l}}\newcommand{\ea}{\end{array}}
\begin{document}
\begin{titlepage}
\nopagebreak
\begin{flushright}

UT-645\\
                June,~1993

\end{flushright}

\vfill
\begin{center}
{\large\bf
Mass formulae of 4-dimensional dilaton black holes}

\vfill
{\bf Tadashi Okai}\\
\it
Department of Physics, Univerity of Tokyo\\
Bunkyo-ku, Tokyo 113, Japan
\rm
\\

\end{center}
\vfill

\begin{abstract}
Integral and differential mass formulae of 4-dimensional stationary
and axisymmetric Einstein-Maxwell-dilaton systems are derived.
The total mass (energy) of these systems are expressed in terms of
other physical quantities such as electric charge of the black hole
suitably modified due to the existence of the dilaton field.
It is shown
that when we vary slightly the fields (metric of the spacetime
$g_{\mu\nu}$, $U(1)-$gauge potential $A_{\mu}$, and dilaton $\phi$) in
such a way as they obey classical equations of motion, the variation
of the dilaton does not contribute explicitly to the variation of the
total mass, but contributes only through the variation of the
electric charge of the black hole.
\end{abstract}
\vfill

\end{titlepage}

\section{Introduction}
One of the great achievements of black hole physics in late
1960's and 1970's,
which have formed the standard picture
of black holes\cite{h-ellis},
was the area theorem proved by Hawking. It has
consequently stimulated the investigations on the close analogy
between black hole mechanics and thermodynamics\cite{beken}\cite{car}.
They revealed the
parallelism between the area of the horizon and the entropy, the
surface gravity and the temperature. Furthermore,
Hawking made it clear that the surface gravity of the horizon is
proportional to the temperature of blackbody radiation of black holes,
induced by their strong gravitational force\cite{therm}.

The study of string theory, on the other hand, offered a new insight
into gravity: the Einstein equation is a low-energy
approximation in string theory\cite{gsw}. New types of actions
including gravity were then considered in the study of low
energy phenomenology of string theory and in the study of string
compactification (including Kaluza-Klein's)\cite{string}.
In recent years, several stringy black hole solutions were discovered
and their physical properties were
examined\cite{maeda}\cite{garf}\cite{limit}.

In this article, we examine the relation between total mass and
other physical quantities in the Einstein-Maxwell-dilaton black holes.
The similarity of black hole mechanics and thermodynamics established
in early 1970's is re-investigated in the context of stringy black
holes. The differential mass formula for Einstein-Maxwell theory was
given by
\bea
\delta M=\Phi_{H}\delta Q_{H}+\Omega_{H}\delta J+T\delta A, \nonumber
\eea
where $\Phi_{H}$, $Q_{H}$, $\Omega_{H}$, $J$, $T$, $A$ denote electric
potential
of electromagnetic gauge field measured at the horizon, electric
charge of the black hole, angular velocity of the horizon,
total angular momentum of the system, Hawking temperature ,
area of the horizon, respectively. (They will be precisely
defined in subsequent sections.)
When we look at the action of Einstein-Maxwell-dilaton theory
\bea
S=\int d^{4}x\sqrt{-g}(R-2(\partial\phi )^{2}-e^{-2\phi }F^{2}),
\label{action}
\eea
the existence of the dilaton field seems, at first sight, to add
extra degrees of freedom to the variation of the total mass.
We will show, however, that this
does not happen
and the effect of the dilation is absorbed into a suitable
redefinition of $Q_{H}$. This result is consistent with the
{\lq\lq}no-hair conjecture\cite{nohair}" in stringy black holes.

The organisation of this article is the following.
In section 2, we recall general setups for the description of
stationary and axisymmetric spacetimes and derive the integral mass
formula of Einstein-Maxwell-dilaton theory. In section 3, we obtain
the differential mass formula. The basic strategy is to use special
properties of Killing vectors. Section 4 is devoted to conclusions and
discussions.

\section{Integral mass formula of dilaton black holes}
We begin by the action (\ref{action}). The equations of motion are
\bea
G_{\mu\nu}-T^{(F)}_{\mu\nu}-T^{(dil)}_{\mu\nu}=0, \;\;\;
\nabla_{\mu}(e^{-2\phi }F^{\mu\nu})=0,\;\;\;
\Box\phi +\frac{1}{2}e^{-2\phi} F^{2}=0, \nonumber
\eea
where we put
\bea
T^{(F)}_{\mu\nu}:=e^{-2\phi }(2F_{\mu \alpha }F_{\nu}^{~~\alpha }
-\frac{1}{2}F^{2}g_{\mu\nu}),\;\;\;
T^{(dil)}_{\mu\nu}:=(2\partial_{\mu}\phi\partial_{\nu}\phi
-(\partial\phi )^{2}g_{\mu\nu}).
\nonumber
\eea

We consider stationary and axisymmetric systems throughout this
article. Furthermore, we restrict ourselves to the systems of
{\lq\lq}rotating bodies." That is to say, the spacetime is invariant
under the simultaneous inversion of time $t$ and azimuthal angle
$\phi$. By the existence of a regular horizon and the gauge freedom
of the metric, we may suppose, without loss of generality, that the
metric be
\bea
ds^{2}=-\sqrt{\Delta(r)}f(r, \theta )
\left(\chi  dt^{2}-\frac{1}{\chi }
(d\phi-\omega dt)^{2}\right)
+g(r, \theta )\left(\frac{dr^{2}}{\Delta(r)}+d\theta ^{2}\right),
\nonumber
\eea
where $f$, $g$, $\chi$, $\omega$ are some functions of $(r, \theta )$
and $\Delta$ of $r$ only (See, e.g., \cite{wald}\cite{chandra}).
We do not need detailed information of $f$, $g$, $\chi$, $\omega$.
Important is that
the horizon is characterised by $\Delta(r)=0$ and the determinant of
the metric of the 2-dimensional space spanned by the vectors
$\partial_{t}$ and $\partial_{\phi}$ is
$g_{tt}g_{\phi\phi}-g_{t\phi}^{2}=-f^{2}\Delta$. Therefore
\bea
g_{tt}g_{\phi\phi}-g_{t\phi}^{2}&\longrightarrow& 0 \nonumber\\
g_{rr} \;\;\;\;\;\;&\longrightarrow& \infty \nonumber
\eea
as one approaches the horizon, whereas
$(g_{tt}g_{\phi\phi}-g_{t\phi}^{2})g_{rr}$ remains finite.
This property is basic in proving the lemma 3 in section 3.

Let $\Sigma$ be a 3-dimensional submanifold of the spacetime defined
by a time-slice of $\{ t=\mbox{constant} \}$. $\Sigma$ has
its spatial infinity, which we denote symbolically by $\infty$, and
is bounded by the outer horizon $H$:
\bea
\partial\Sigma=\{\infty   \}\cup \{ H \}. \nonumber
\eea

Let us recall now several physical quantities needed to develop
integral and differential mass formulae\cite{car}\cite{wald}. Let
\bea
k^{\cdot}:=\left( \frac{\partial }{\partial t} \right), \;\;\;
m^{\cdot}:=\left( \frac{\partial }{\partial\phi } \right) \nonumber
\eea
be the time translational killing vector and axisymmetric Killing
vector, respectively.
The angular velocity of the black hole and the electric potential
measured at the horizon are defined by
\bea
\Omega_{H}:=-\left[\frac{g_{t \phi }}{g_{\phi\phi }} \right]_{H}
\;\;\; \mbox{and } \;\;\;
\Phi_{H}:=\left[ l^{\mu}A_{\mu} \right]_{H},
\mbox{respectively.} \nonumber
\eea
Since we are now assuming the spacetime to be axisymmetric, so are the
spatial infinity $\infty$ and the outer horizon $H$. This means that
the axial Killing vector $m^{\cdot}$
is tangent to $\infty$, $H$, and $\Sigma$. Thus
$m^{\mu}d\sigma _{\mu}=0,\;\;\;$ $m^{\mu}dS_{\mu\nu}=0$
( both on $\infty$ and $H$ ), where $d\sigma_{\mu}$ and $dS_{\mu\nu}$
are the dual of volume elements of $\Sigma$ and $\partial\Sigma$,
respectively.

Another Killing vector, which is null on $H$ and is normal to $H$,
is defined as follows:
\bea
l^{\cdot}:=\left( \frac{\partial }{\partial t} \right)
+\Omega_{H} \left( \frac{\partial }{\partial\phi } \right).
\nonumber
\eea
(Note that $l^{\cdot}$ is tangent to the 3-dimensional space that the
2-dimensional space $H$
sweeps with time evolution.)
The other null vector orthogonal to $H$ is denoted by $n^{\cdot}$,
which is normalised such that $n_{\mu}l^{\mu}=-1$. Then the dual of
the surface element of $\partial\Sigma$ is expressed as
\bea
dS_{\mu\nu}=\frac{1}{2}(l_{\mu} n_{\nu}-n_{\mu} l_{\nu})dA
=l_{[\mu}n_{\nu]}dA, \nonumber
\eea
where $dA$ is the surface area element of $\partial\Sigma$.
We use semicolon {\lq\lq};" in the meaning of the covariant
differentiation $\nabla_{\mu}$ throughout this paper.

The total mass $M$ of the system, the total angular momentum $J$, and
the total {\lq\lq}modified" electric charge $\tilde{Q}$ are defined
by the surface integral at spatial infinity:
\bea
M:=-\frac{1}{4\pi}\int_{\infty}k^{\mu;\nu}dS_{\mu\nu},\;\;\;
J:=-\frac{1}{8\pi}\int _{\infty} m^{\mu;\nu}dS_{\mu\nu},\;\;\;
\tilde{Q}
:=-\frac{1}{4\pi}\int_{\infty}e^{-2\phi}F^{\mu\nu}dS_{\mu\nu}.
\nonumber
\eea
These quantities are expressed in other ways:
\bea
M=\frac{1}{4\pi}\int _{H}k^{\alpha;\beta }dS_{\alpha\beta }
-\frac{1}{4\pi}\int  (T^{\alpha\beta }-\frac{1}{2}
Tg^{\alpha\beta })k_{\beta }d\sigma_{\alpha}.
\label{mass}
\eea
Here we have used Stokes' theorem
\bea
\int V^{\mu\nu}_{~~~~;\nu}d\sigma _{\mu}
=\left( \int _{\infty}+\int _{H} \right)V^{\mu\nu}dS_{\mu\nu}
\;\;\;\mbox{for all $V^{\mu\nu}=V^{[\mu\nu]}$} \nonumber
\eea
and the equation of motion
$R_{\mu\nu}=T_{\mu\nu}-\frac{1}{2}g_{\mu\nu}T$. Note that
any Killing vector $\xi^{\mu}$ satisfies the equality
$\xi^{\mu;\alpha}_{~~~;\alpha}=-R^{\mu}_{~~~\nu}\xi^{\nu}$.
Similarly, total angular momentum and modified electric charge are
rewritten as
\bea
J&=&\frac{1}{8\pi}\int_{H}m^{\mu;\nu}dS_{\mu\nu}
+\frac{1}{8\pi}\int T^{(F)}_{\mu\nu}m^{\mu}d\sigma^{\nu}
=:J_{H}+J_{F}, \nonumber \\
\tilde{Q}
&=&\frac{1}{4\pi}\int_{H}e^{-2\phi}F^{\mu\nu}dS_{\mu\nu}
=:\tilde{Q}_{H}. \nonumber
\eea
We naturally assume the stationarity and axisymmetry of dilaton
and electromagnetic field as well as the metric. Thus
\bea
({\cal L}_{k}g)_{\mu\nu}=({\cal L}_{k}\phi )=
({\cal L}_{k}A)_{\mu}=0, \;\;\;
({\cal L}_{m}g)_{\mu\nu}=({\cal L}_{m}\phi )=
({\cal L}_{m}A)_{\mu}=0, \nonumber
\eea
where ${\cal L}$ denotes the Lie derivative.
Therefore
\bea
T^{(dil)}_{\mu\nu}m^{\mu}d\sigma^{\nu}=\left( 2\phi_{;\mu}\phi_{;\nu}
-(\partial\phi)^{2}g_{\mu\nu}\right) m^{\mu}d\sigma^{\nu} \nonumber
\eea
vanishes.
Thus there is no contribution of dilaton to the total angular
momentum.
And we note here that surface gravity $\kappa$ and electric potential
$\Phi_{H}$ are constants on the horizon if the dominant energy
condition of the energy momentum tensor is satisfied. It is naturally
satisfied in our present case.
We do no explain details of these quantities.
The reader is referred to \cite{car}\cite{wald} for the discussion of
their physical meaning and property.

First, the following lemma is a starting point of the integral mass
formula.
\\
\\
{\bf lemma 1}\\
Let $A$ be the area of the event horizon of the black hole and
$T:=\kappa/(8\pi)$ be the Hawking temperature of the horizon. Then the
total mass (energy) of the system is given by
\bea
\frac{M}{2}=\int T_{\alpha\beta}k^{\alpha}d\sigma^{\beta}
+\frac{1}{16\pi}\int R
(k^{\mu}d\sigma_{\mu})+\Omega_{H} J_{H}+TA. \nonumber
\eea
{\bf proof}\\
Using Stokes' theorem,
\bea
\frac{M}{2}=-\frac{1}{8\pi} \int_{\infty}
k^{\alpha;\beta}dS_{\alpha\beta}
=-\frac{1}{8\pi}\int k^{\alpha;\beta}_{~~~~;\beta}d\sigma_{\alpha}
+\frac{1}{8\pi}\int_{H}k^{\alpha;\beta}dS_{\alpha\beta}. \nonumber
\eea
Next, using the property of Killing vectors
$k^{\alpha;\beta}_{~~~~;\beta}=-R^{\alpha}_{~~\beta}k^{\beta}$,
\bea
\frac{M}{2}&=&\frac{1}{8\pi} \int
R^{\alpha}_{~~\beta}k^{\beta}d\sigma_{\alpha}
+\frac{1}{8\pi} \int_{H}(l^{\alpha;\beta}-\Omega_{H} m^{\alpha;\beta})
dS_{\alpha\beta} \nonumber\\
&=&\frac{1}{8\pi} \int T_{\alpha\beta}k^{\alpha}d\sigma^{\beta}
+\frac{1}{16\pi}\int
R(k^{\alpha}d\sigma_{\alpha})+TA+\Omega_{H} J_{H}.
\;\;\;\mbox{Proof ends.} \nonumber
\eea
This lemma is followed by the proposition 1, which states the integral
mass formula of Einstein-Maxwell-dilaton theory.
\\
\\
{\bf proposition 1}\\
Under these assumptions, total mass of the system $M$ is rephrased as
\bea
\frac{M}{2}=TA+\Omega_{H}(J_{H}+J_{F})+\tilde{Q}_{H} \Phi_{H}
+\frac{1}{16\pi}\int (R-2(\partial \phi )^{2}-e^{-2\phi } F^{2})
(k^{\mu}d\sigma_{\mu}).
\nonumber
\eea
{\bf proof}\\
Our claim is
\bea
\frac{1}{8\pi}\int T_{\alpha \beta }k^{\alpha }d\sigma^{\beta }=
\Phi_{H} \tilde{Q}_{H} +\Omega_{H} J_{F}
-\frac{1}{16\pi}\int (2(\partial \phi )^{2}+e^{-2\phi } F^{2})
(kd\sigma).
\nonumber
\eea
First,
\bea
\frac{1}{8\pi}\int T_{\alpha \beta }k^{\alpha }d\sigma^{\beta }
&=&\frac{1}{8\pi}\int e^{-2\phi }
(2F_{\alpha \mu}F_{\beta }^{~~\mu}-\frac{1}{2}F^{2}g_{\mu\nu})
(l^{\alpha }-\Omega_{H} m^{\alpha })d\sigma^{\beta } \nonumber \\
&&+\frac{1}{8\pi}\int
(2\phi _{;\alpha }\phi _{;\beta }-(\partial \phi )^{2}g_{\mu\nu})
k^{\alpha }d\sigma^{\beta } \nonumber \\
&=&\frac{1}{4\pi}\int e^{-2\phi } F_{\alpha \mu}F_{\beta }^{~~\mu}
l^{\alpha }d\sigma^{\beta }
+\Omega_{H} J_{F} \nonumber \\
&&-\frac{1}{16\pi}\int
(2(\partial \phi )^{2}+e^{-2\phi } F^{2})(kd\sigma).
\nonumber
\eea
Thus, if we prove
\bea
\frac{1}{4\pi}\int e^{-2\phi }
F_{\alpha \mu}F_{\beta }^{~~\mu}l^{\alpha }d\sigma^{\beta }
=\Phi_{H} \tilde{Q}_{H},
\nonumber
\eea
then the proof of the proposition will be completed.
{}From the Killing symmetry of $A_{\mu}$, we have
$F_{\alpha \mu}F_{\beta }^{~~\mu}l^{\alpha }
=l^{\alpha }(A_{\mu;\alpha }-A_{\alpha ;\mu})F_{\beta }^{~~\mu}
=-(l^{\alpha }A_{\alpha })_{;\mu}F_{\beta }^{~~\mu}$.
Therefore, using the equation of motion of $A_{\mu}$,
\bea
\frac{1}{4\pi}\int e^{-2\phi } F_{\alpha \mu}F_{\beta }^{~~\mu}
l^{\alpha }d\sigma^{\beta }
&=&-\frac{1}{4\pi}\int e^{-2\phi }
(l^{\alpha }A_{\alpha })_{;\mu}F^{\beta \mu}d\sigma_{\beta }
\nonumber \\
&=&-\frac{1}{4\pi}\int (e^{-2\phi }
(l^{\alpha }A_{\alpha })F^{\beta \mu})_{;\mu}d\sigma_{\beta }
\nonumber \\
&=&\frac{1}{4\pi}\int _{H}e^{-2\phi } (l^{\alpha }A_{\alpha })
F^{\beta \mu}dS_{\beta \mu} \nonumber \\
&=&\Phi_{H} \tilde{Q}_{H}.\;\;\; \mbox{Proof ends.}\nonumber
\eea

Let us check that the electric potential is constant on $H$ in the
Einstein-Maxwell-dilaton theory as well as in the Einstein-Maxwell.
Actually, $R_{\mu\nu}l^{\mu}l^{\nu}=0$ holds on $H$. Using the
equation of motion, we get
$(T^{(F)}_{\mu\nu}+T^{(dil)}_{\mu\nu})l^{\mu}l^{\nu}=0$.
Now
\bea
T^{(dil)}_{\mu\nu} l^{\mu}l^{\nu}
=(2\partial_{\mu}\phi\partial_{\nu}\phi-g_{\mu\nu}(\partial\phi)^{2})
l^{\mu}l^{\nu}=0 \nonumber
\eea
is easily seen from the Killing symmetry. Thus
$T^{(F)}_{\mu\nu}l^{\mu}l^{\nu}=0.$ The rest of the proof is
exactly the same as that of the Einstein-Maxwell theory. Therefore
$\Phi_{H}$ is constant even when dilaton is coupled.

We note here that $J_{F}$ is rewritten, by using
$F_{\alpha \mu}F_{\beta }^{~~\mu}m^{\alpha }
=m^{\alpha }(A_{\mu;\alpha }-A_{\alpha ;\mu})F_{\beta }^{~~\mu}
=-(m^{\alpha }A_{\alpha })_{;\mu}F_{\beta }^{~~\mu}$, in the following
way:
\bea
J_{F}=\frac{1}{4\pi}\int_{H}e^{-2\phi}
(m^{\mu}A_{\mu})F^{\alpha\beta}dS_{\alpha\beta}. \nonumber
\eea

\section{Differential mass formula}
In this section, we carry out the variational calculation
and show the differential mass formula, which is
stated in the proposition 2.
When we vary black hole solutions,
we preserve the horizon, stationarity and axisymmetry of the black
hole\cite{car}:\\
\it
\\
1)Since we assume the spacetime be stationary and axisymmetric,
we should preserve these symmetries before and after the variation.\\
2)There is a freedom of the general coordinate transformation.
Using this freedom, we can require that the horizon $H$ itself be
invariant before and after the variation and that the null vector $l$
remain normal to $H$. \\
\\
\rm
Thus,
$\delta k^{\mu}=\delta m^{\mu}=0$,
$\delta l^{\mu}=\delta \Omega_{H}m^{\mu}$, and
$(\delta l)^{[\mu} l^{\nu]}=0$ holds.
These conditions are important in order to prove the lemma 2.

{\bf proposition 2}
\bea
\delta M=T\delta A+\Omega_{H}(\delta J_{H}+\delta J_{F})
+\Phi_{H} \delta\tilde{Q}_{H}.
\label{alpha}
\eea
We provide some lemmas below in order to prove the proposition 2.\\
\\
{\bf lemma 2}
\bea
&&\delta\left( \frac{1}{16\pi}\int R(k^{\mu}d\sigma_{\mu}) \right)
\nonumber \\
&&~~~~~~~~=-\frac{1}{16\pi }\int G^{\mu\nu}\delta
g_{\mu\nu}(k^{\alpha}d\sigma_{\alpha})
-\frac{1}{4\pi }\left(\int_{\infty}+\int_{H}\right)\delta
g_{\mu}^{~~[\mu;\beta]}
k^{\alpha}dS_{\alpha\beta} \nonumber\\
&&~~~~~~~~=-\frac{1}{16\pi }\int G^{\mu\nu}\delta
g_{\mu\nu}(k^{\alpha}d\sigma_{\alpha})
-\frac{1}{2}\delta M-A\delta T-J_{H}\delta\Omega_{H}.
\label{beta}
\eea
The proof of the lemma 2 is given in \cite{car}.
\\
\\
{\bf lemma 3}\\
When the Einstein-Maxwell-dilaton system is stationary and
axisymmetric, and the spacetime has a regular horizon, the following
equality holds:
\bea
\int_{H}l^{\mu}\delta A_{\mu}F^{\alpha\beta}dS_{\alpha\beta}=-2
\int_{H}\delta A_{\mu}F^{\mu\alpha}l^{\beta}dS_{\alpha\beta}. \nonumber
\eea
{\bf proof of lemma 3}\\
According to the assumption of the symmetry, we can assume without
loss of generality that
\bea
A=A_{\mu}dx^{\mu}=A_{t}dt+A_{\phi }d\phi\nonumber
\eea
and that the dual of the surface element $dS_{\mu\nu}$ has the only
component $dS_{tr}$.
Then,
\bea
\delta A_{\mu}F^{\mu\alpha}l^{\beta}dS_{\alpha\beta}&=&
(\delta A_{\phi }F^{\phi t}l^{r}-\delta A_{t}F^{tr}l^{t}-\delta A_{\phi
}F^{\phi r}
l^{t})dS_{tr} \nonumber\\
&=&(\delta A_{t}F^{tr}-\delta A_{\phi }F^{\phi r})dS_{tr} \nonumber
\eea
and
\bea
\delta A_{\mu}l^{\mu}F^{\alpha\beta}dS_{\alpha\beta}=
2(\delta A_{t}+\Omega_{H}\delta A_{\phi })F^{tr}dS_{tr}. \nonumber
\eea
Since we have assumed the spacetime to have regular horizons,
$-g_{tt}g_{\phi\phi}+(g_{t\phi})^{2}$ goes to zero and $g_{rr}$ goes
to infinity with
$(-g_{tt}g_{\phi\phi}+(g_{t\phi})^{2})/g_{rr}$ remaining finite as we
approach the horizon. Therefore
we can calculate on the horizon
\bea
g^{rr}\Omega_{H}g^{tt}
&=&-g^{rr}\frac{g_{t\phi}}{g_{\phi\phi}}
\cdot\frac{g_{\phi\phi}}{g_{tt}g_{\phi\phi}-(g_{t\phi})^{2}}
=\frac{-g^{rr}g_{t\phi}}{g_{tt}g_{\phi\phi}-(g_{t\phi})^{2}}
=g^{rr}g^{t\phi}, \nonumber\\
g^{rr}\Omega_{H}g^{\phi t}&=&
-g^{rr}\frac{g_{t\phi}}{g_{\phi\phi}}
\cdot\frac{-g_{t\phi}}{g_{tt}g_{\phi\phi}-(g_{t\phi})^{2}}
=\frac{g^{rr}g_{tt}}{g_{tt}g_{\phi\phi}-(g_{t\phi})^{2}}
=g^{rr}g^{\phi\phi}. \nonumber
\eea
Thus we get
\bea
\left[\Omega_{H}F^{tr}\right]_{H}
&=&\Omega_{H}\left[g^{rr}(g^{tt}F_{rt}+g^{\phi t}
F_{\phi r}) \right]_{H} \nonumber\\
&=&\left[g^{rr}(g^{t\phi }F_{tr}-g^{\phi \phi }F_{\phi r})\right]_{H}
=\left[F^{\phi r}\right]_{H}. \nonumber
\eea
Therefore
\bea
\delta A_{\mu}l^{\mu}F^{\alpha\beta}dS_{\alpha\beta}&=&
2(\delta A_{t}+\Omega_{H}\delta A_{\phi })F^{tr}dS_{tr}
=2(\delta A_{t}F^{tr}+\delta A_{\phi }F^{\phi r})dS_{tr} \nonumber \\
&=&-2\delta A_{\mu}F^{\mu\alpha}l^{\beta}dS_{\alpha\beta} \nonumber
\eea
holds on the horizon. This proves the lemma 3.
\\
{\bf lemma 4}
\bea
&&\delta\left(-\frac{1}{16\pi }\int e^{-2\phi }
F^{2}(k^{\mu}d\sigma_{\mu}) \right) \nonumber \\
&&~~~~~=-\frac{1}{16\pi }\int (-e^{-2\phi })
\left(2F^{\alpha\mu}F_{\alpha}^{~~\nu}-
\frac{1}{2}F^{2}g^{\mu\nu}\right)\delta g_{\mu\nu}
(k^{\alpha}d\sigma_{\alpha}) \nonumber\\
&&~~~~~+\frac{1}{8\pi }\int \delta\phi e^{-2\phi } F^{2}(k^{\mu}d\sigma_{\mu})
-\tilde{Q}_{H} \delta\Phi_{H}-J_{F} \delta\Omega_{H}.
\label{gamma}
\eea
{\bf proof of lemma 4}\\
Let
$\delta_{g}:=g_{\mu\nu}$-variation,
$\delta_{A}:=A_{\mu}$-variation, and
$\delta_{\phi}:=\phi$-variation. Then
\bea
\delta_{g}\left(-\frac{1}{16\pi }\int e^{-2\phi }F^{2}(k^{\mu}d\sigma_{\mu})
\right)
&=&-\frac{1}{16\pi }\int e^{-2\phi }(2F^{\mu\alpha}F^{\nu}_{~~\alpha}
-\frac{1}{2}F^{2}g^{\mu\nu})
\delta g_{\mu\nu}(k^{\beta}d\sigma_{\beta}), \nonumber\\
\delta_{\phi }\left(-\frac{1}{16\pi }\int e^{-2\phi }
F^{2}(k^{\mu}d\sigma_{\mu}) \right)
&=&-\frac{1}{8\pi }\int \delta\phi e^{-2\phi }F^{2}(k^{\mu}d\sigma_{\mu})
\nonumber
\eea
is easily obtained. The proof of the lemma is completed if we prove
\bea
(\#):=
\delta _{A}\left(-\frac{1}{16\pi }\int e^{-2\phi } F^{2}(k^{\mu}d\sigma_{\mu})
\right)
=-\tilde{Q}_{H} \delta\Phi_{H}-J_{F} \delta\Omega_{H}. \nonumber
\eea
{}From the equation of motion of $A_{\mu}$ and the Killing
symmetry of the system, we have
\bea
e^{-2\phi}F^{\alpha\beta}l^{\mu}
&=&(e^{-2\phi }F^{\alpha\beta}\delta A_{\beta}l^{\mu})_{;\alpha}
-e^{-2\phi }F^{\alpha\beta}\delta A_{\beta}l^{\mu}_{~~;\alpha}
\nonumber\\
&=&(e^{-2\phi }F^{\alpha\beta}\delta A_{\beta}l^{\mu})_{;\alpha}
-l^{\alpha}(e^{-2\phi }F^{\mu\beta}\delta A_{\beta})_{;\alpha}
\nonumber\\
&=&(e^{-2\phi }F^{\alpha\beta}\delta A_{\beta}l^{\mu})_{;\alpha}
-(e^{-2\phi }F^{\mu\beta}\delta A_{\beta}l^{\alpha})_{;\alpha}.
\nonumber
\eea
Thus, using this formula,
\bea
(\#)&=&-\frac{1}{4\pi}\int
e-2\phi F^{\alpha\beta}\delta A_{\beta;\alpha}(ld\sigma)
=-\frac{1}{2\pi}\int_{H}e^{-2\phi }
F^{\alpha\beta}\delta A_{\beta}l^{\mu}dS_{\mu\alpha}
\nonumber\\
&=&\frac{1}{4\pi }\int_{H}
e^{-2\phi }F^{\alpha\beta}\delta A_{\mu}l^{\mu}dS_{\alpha\beta},
\nonumber
\eea
here we have used the lemma 3. Applying
\bea
\left[l^{\mu}\delta A_{\mu}\right]_{H}
=\delta\Phi_{H}-(m^{\mu}A_{\mu})\delta\Omega_{H}
\;\;\; \mbox{and} \;\;\;
J_{F}=\frac{1}{4\pi}\int_{H}e^{-2\phi}
(m^{\mu}A_{\mu})F^{\alpha\beta}dS_{\alpha\beta}
\nonumber
\eea
to the above, we get
\bea
(\#)=-\tilde{Q}_{H} \delta\Phi_{H}-J_{F}\delta\Omega_{H}. \nonumber
\eea
We have completed the proof of lemma 4.
\\
{\bf lemma 5}
\bea
&& \delta\left(-\frac{1}{16\pi}\int
2(\partial\phi )^{2}(kd\sigma)\right) \nonumber\\
&&=-\frac{1}{16\pi}\int (-2\phi^{;\mu}\phi^{;\nu}
+(\partial \phi )^{2}g^{\mu\nu})\delta g_{\mu\nu}
(kd\sigma)+\frac{1}{4\pi}\int\delta\phi(\Box \phi )(kd\sigma).
\label{delta}
\eea
{\bf proof of lemma 5}\\
\bea
\delta\left(-\frac{1}{16\pi}\int
2(\partial\phi )^{2}(kd\sigma)\right)
&=&(\delta _{g}+\delta _{\phi })
\left(-\frac{1}{16\pi}\int 2(\partial\phi )^{2}
(kd\sigma)\right) \nonumber\\
&=&-\frac{1}{16\pi}\int((\partial\phi)^{2}g^{\mu\nu}
-2\phi^{;\mu}\phi^{;\nu} )\delta g_{\mu\nu}(kd\sigma) \nonumber\\
&&-\frac{1}{4\pi}\int(\delta\phi)_{;\mu}\phi^{;\mu}(kd\sigma).
\label{deldil}
\eea
Now we can use the Killing symmetry and we get
\bea
(\delta\phi )_{;\mu}\phi ^{;\mu}k^{\alpha}&=&(\delta\phi\phi
^{;\mu}k^{\alpha})_{;\mu}
-\delta\phi(\Box \phi )k^{\alpha}-\delta\phi\phi ^{\mu}k^{\alpha}_{~~~;\mu}
\nonumber\\
&=&(\delta\phi\phi ^{;\mu}k^{\alpha})_{;\mu}
-\delta\phi(\Box \phi )k^{\alpha}-k^{\mu}(\delta\phi\phi ^{;\alpha})_{;\alpha}
\nonumber\\
&=&(\delta\phi\phi ^{;\mu}k^{\alpha})_{;\mu}
-(\delta\phi\phi ^{;\alpha}k^{\mu})_{;\mu}
-\delta\phi(\Box \phi )k^{\alpha}. \nonumber
\eea
Plugging this equality into eq.(\ref{deldil}), we obtain
\bea
-\frac{1}{4\pi}\int(\delta\phi)_{;\mu}\phi^{;\mu}(kd\sigma)
&=&-\frac{1}{2\pi}\int(\delta\phi)\phi^{;\mu}k^{\alpha}
dS_{\alpha\mu}
+\frac{1}{4\pi}\int\delta\phi (\Box \phi ) (kd\sigma) \nonumber\\
&=&-\frac{1}{4\pi}\int(\delta\phi)\phi^{;\mu}k^{\alpha}(l_{\mu}n_{\nu}-n_{\mu}l_{\nu})
dA+\frac{1}{4\pi}\int\delta\phi (\Box \phi ) (kd\sigma) \nonumber\\
&=&\frac{1}{4\pi}\int\delta\phi (\Box \phi ) (kd\sigma). \nonumber
\eea
This proves the lemma 5.

Putting the eq.s(\ref{beta}), (\ref{gamma}), and (\ref{delta})
together, we obtain the differential mass formula (\ref{alpha}).

\section{Conclusion and discussion}
We presented the integral and differential mass formulae of
Einstein-Maxwell-dilaton system. Electric charge usually defined
in Einstein-Maxwell system
$Q_{H}:=\frac{1}{4\pi}\int_{H}F^{\mu\nu}dS_{\mu\nu}$
is now replaced by $\tilde{Q}_{H}$. This is quite reasonable when
we look
at the equation of motion of the electromagnetic field. Interesting
is that the only term, in the differential mass formula, explicitly
related to the variation of dilaton is $\delta  \tilde{Q}_{H}$.
Dilaton in the
Einstein-Maxwell-dilaton theory is an
extra degree of freedom in the action, compared with the
Eintein-Maxwell theory. It does not give, however, any further
contribution to the globally defined quantities such as mass, charge, or
angular momentum.

We conventionally interpret the 1st term of the right hand side of
the eq.(\ref{mass}) as
the mass of the black hole and the 2nd term as the
contribution to the total mass of the matter (electromagnetic field
and dilaton field in the present case) outside the horizon\cite{car}.
Now let us
consider an extremally charged spherical black hole
configuration\cite{garf}\cite{limit}:
\bea
ds^{2}&=&-(1-2M/r)dt^{2}
+\frac{dr^{2}}{(1-2M/r)}+(1-2M/r)d\Omega, \nonumber\\
A_{\mu}dx^{\mu}&=&\sqrt{2}Mdt/r, \;\;\; e^{2\phi}=(1-2M/r). \nonumber
\eea
Clearly, $\int _{H}k^{\alpha;\beta }dS_{\alpha\beta }=0$ holds when
the
black hole is
extremal. If we take it literally, the mass of the black hole is zero.
But, we should note that this zero mainly comes from the fact that
the area of the horizon vanishes as the black hole approaches its
extremality.
The total mass of the system is defined, in one way,
by using a virtual 2-dimensional surface and test body of unit mass
placed on it
(See, e.g., \cite{wald}). This method clearly breaks down when the area
of the 2-dimensional surface under consideration shrinks to zero.
On the other hand, the modified electric charge
$\tilde{Q}_{H}$
is nonzero and finite even when the area shrinks to zero.
This suggests us that the modification of the definition of
mass might be needed as well as that of electric charge. We leave this
problem for future investigation.
\\
\\
\\
\footnotesize {\it Acknowledgments.}
It is of great pleasure to express my gratitude to
Prof.T.Eguchi
for sincere advice, guidance, and valuable comments. I would like to
acknowledge Prof.J.Arafune for fruitful discussion and encouragement.
I am indebted to Prof.A.Kato and all the members of the
Elementary particle theory group in the University of Tokyo for
support and encouragement.


\normalsize

\begin{thebibliography}{99}
\bibitem{h-ellis}S.W.Hawking and G.F.R.Ellis, {\lq\lq}The large scale
	structure of space-time," Cambridge University Press, 1973
\bibitem{beken}J.Bekenstein, Phys. Rev. D7 (1973) 2333; \\
	J.Bekenstein, Phys. Rev. D12 (1975) 3077
\bibitem{car}B.Carter, {\lq\lq}Black Hole Equilibrium States II,"
	in {\it Black Holes}, ed. C.DeWitt and B.S.DeWitt ( New York:
	Gordon \& Breach, Les Houses lecture notes );\\
	J.M.Bardeen, B.Carter, and S.W.Hawking, Comm. Math. Phys. 31 (1973) 161
\bibitem{therm} S.W.Hawking, Phys. Rev. 13 (1976) 191;\\
	J.B.Hartle and S.W.Hawking, Phys. Rev. D13 (1976) 2188
\bibitem{gsw}M.B.Green, J.H.Schwarz, and E.Witten, {\lq\lq}Superstring
	Theory," Cambridge University Press, 1987
\bibitem{string}C.G.Callan, J.A.Harvey, and A.Strominger, Nucl. Phys.
	B359 (1991) 611;\\
	C.G.Callan, D.Friedan, E.J.Martinec, and M.J.Perry, Nucl. Phys.
	B262 (1985) 593;\\
	C.G.Callan, R.C.Myers, and M.J.Perry,
	Nucl. Phys. B311 (1988/1989) 673;\\
	D.J.Gross and J.H.Sloan, Nucl. Phys. B291 (1987) 41
\bibitem{maeda} G.W.Gibbons and K.Maeda, Nucl. Phys. B298 (1988) 741
\bibitem{garf} D.Garfinkle, G.T.Horowitz, and A.Strominger,
	Phys. Rev. D43 (1991) 3140; \\
	G.T.Horowitz, {\lq\lq}The Dark Side of String Theory: Black
	Holes and Black Strings," Santa Barbara preprint, UCSBTH-92-32,
	hep-th/9210119, and the references therein
\bibitem{limit}J.Preskill, A.Shapere, S.Trivedi, and F.Wilczek,
	Mod. Phys. Lett. A6 (1991) 2353; \\
	C.F.E.Holzhey and F.Wilczek, Nucl. Phys. B380 (1992) 447
\bibitem{nohair}W.Israel, Phys. Rev. 25 (1967) 1776;\\
	D.C.Robinson, Phys. Rev. 10 (1974) 458;\\
	D.C.Robinson, Phys. Rev. Lett. 34 (1975) 905;\\
	B.Carter, Phys. Rev. Lett. 26 (1971) 331
\bibitem{wald} R.M.Wald, {\lq\lq}General relativity," the University of
	Chicago press, 1984
\bibitem{chandra} S.Chandrasekhar, {\lq\lq}Mathematical theory of black
	holes," Oxford University press, 1983
\end{thebibliography}
\end{document}